# MUON INTENSITY INCREASE BY WEDGE ABSORBERS FOR LOW-E MUON EXPERIMENTS*

D. V. Neuffer#, D. Stratakis, J. Bradley&, Fermilab, Batavia IL 60510, USA

*Abstract*

Low energy muon experiments such as mu2e and g-2 have a limited energy spread acceptance. Following techniques developed in muon cooling studies and the MICE experiment, the number of muons within the desired energy spread can be increased by the matched use of wedge absorbers. More generally, the phase space of muon beams can be manipulated by absorbers in beam transport lines. Applications with simulation results are presented.

## INTRODUCTION

Low energy muon experiments, such as the Fermilab-based mu2e [1, 2] and g-2 experiments, [3] have a limited phase space acceptance for useful muons. The mu2e experiment can only accept a small momentum slice of the incident muon momentum spectrum (Pμ < ~50 MeV/c, see below). The g-2 experiment only accepts a momentum spread of δP = ±0.1% around the design momentum of ~3.1 GeV/c. Methods that can increase the number of muons within the momentum acceptance are desirable.

Similar or complementary constraints occurred in the exploration of ionization cooling for muons. [4] Wedge absorbers are needed to transform the intrinsic transverse cooling effect to include longitudinal cooling, and introduce exchanges between longitudinal and transverse phase space densities. In cooling channels incremental exchanges were developed so that large increases in phase space density could be obtained over a multistage system. In final cooling toward the extreme parameters needed for a high luminosity collider, it was noted that very large exchanges in single wedges were needed. [5,6] In that limit it was noted that a wedge could be treated as an optical element in a transport system and large exchanges can occur with single wedges, [7] which can be used to match the final beam to desired distributions (smaller transverse emittance with larger δp or vice versa). As an example the use of a wedge, and its effects in large exchanges, can be measured in the MICE experiment. [8, 9]

The method can also be adapted for phase space matching into low energy muon experiments, and matched placement of wedges in the beam transport could obtain more muons within experiment acceptances. We note that use of a wedge to reduce δP increases the transverse emittance, and changes the matched optics. Some iterations in beam matching may be needed to increase the number muons accepted.

In this paper we first describe the wedge process and its approximation as a transport element

The parameters of mu2e and g-2 are discussed and potential uses of wedge absorbers and their adaptation to increase acceptance into the experiments are described. Simulations that test these possibilities are presented and the results are discussed.

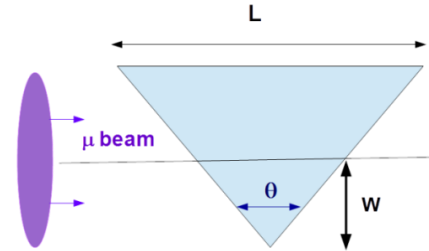

Figure 1: Schematic view of a muon beam passing through a wedge.

## WEDGE EFFECTS ON BEAM - FIRST ORDER ESTIMATES

Figure 1 shows a stylized view of the passage of a beam with dispersion $\eta_0$ through a wedge absorber. The wedge is approximated as an object that changes particle momentum offset $\delta = \Delta p/P_0$ as a function of $x$, and the wedge is shaped such that that change is linear in $x$. (The change in average momentum $P_0$ is ignored in this approximation. Energy straggling and multiple scattering are also ignored.) The rms beam properties entering the wedge are given by the transverse emittance $\varepsilon_0$, betatron amplitude $\beta_0$, dispersion $\eta_0$ and relative momentum width $\delta_0$. (To simplify discussion the beam is focussed to a betatron and dispersion waist at the wedge: $\beta_0'$, $\eta_0' = 0$. This avoids the complication of changes in $\beta'$, $\eta'$ in the wedge.) The wedge is represented by its relative effect on the momentum offsets $\delta$ of particles within the bunch at position $x$:

$$\frac{\Delta p}{p} = \delta \rightarrow \delta - \frac{2(dp/ds)\tan(\theta/2)}{P_0} x = \delta - \delta' x$$

$dp/ds$ is the momentum loss rate in the material ($dp/ds = \beta^{-1}dE/ds$). $2x\tan(\theta/2)$ is the wedge thickness at transverse position $x$ (relative to the central orbit at $x=0$), and $\delta' = 2dp/ds \tan(\theta/2)/P_0$ indicates the change of $\delta$ with $x$.

Under these approximations, the initial dispersion and the wedge can be represented as linear transformations in the $x$-$\delta$ phase space projections and the transformations are phase-space preserving. The dispersion can be represented by the matrix: $\mathbf{M}_\eta = \begin{bmatrix} 1 & \eta_0 \\ 0 & 1 \end{bmatrix}$, since $x \Rightarrow x + \eta_0\delta$. The wedge can be represented by the matrix: $\mathbf{M}_\delta = \begin{bmatrix} 1 & 0 \\ -\delta' & 1 \end{bmatrix}$, obtaining $\mathbf{M}_{\eta\delta} = \begin{bmatrix} 1 & \eta_0 \\ -\delta' & 1-\delta'\eta_0 \end{bmatrix}$. Writing the $x$-$\delta$ beam distribution as a phase-space ellipse:

$g_0 x^2 + b_0 \delta^2 = \sigma_0 \delta_0$, and transforming the ellipse by standard betatron function transport techniques obtains new coefficients $b_1$, $g_1$, $a_1$, which define the new beam parameters[6]. The momentum width is changed to:

$$\delta_1 = \sqrt{g_1 \sigma_0 \delta_0} = \delta_0 \left[ (1-\eta_0 \delta')^2 + \frac{\delta'^2 \sigma_0^2}{\delta_0^2} \right]^{1/2}.$$

The bunch length is unchanged. The longitudinal emittance has therefore changed simply by the ratio of energy-widths, which means that the longitudinal emittance has changed by the factor $\delta_1/\delta_0$. The transverse emittance has changed by the inverse of this factor:

$$\varepsilon_1 = \varepsilon_0 \left[ (1-\eta_0 \delta')^2 + \frac{\delta'^2 \sigma_0^2}{\delta_0^2} \right]^{-1/2}$$

The new values of $(\eta, \beta)$ are:
$$\eta_1 = -\frac{a_1}{g_1} = \frac{\eta_0(1-\eta_0 \delta') - \delta' \frac{\sigma_0^2}{\delta_0^2}}{(1-\eta_0 \delta')^2 + \delta'^2 \frac{\sigma_0^2}{\delta_0^2}}$$
and
$$\beta_1 = \beta_0 \left[ (1-\eta_0 \delta')^2 + \frac{\delta'^2 \sigma_0^2}{\delta_0^2} \right]^{-1/2}.$$

Note that the change in betatron functions implies that the following optics should be correspondingly matched.

A single wedge exchanges emittance between one transverse dimension and longitudinal; the other transverse plane is unaffected. Serial wedges could be used to balance $x$ and $y$ exchanges, or a more complicated coupled geometry could be developed.

Wedge parameters can be arranged to obtain large exchange factors in a single wedge. In final cooling we wish to reduce transverse emittance at the cost of increased longitudinal emittance.

## APPLICATION TO MU2E

The mu2e experiment presents an unusual opportunity to exploit beam-cooling techniques to improve acceptance. The transport from target to detector includes a bent solenoid that produces a dispersion that invites the introduction of a wedge absorber to shape the transmitted energy distribution.

The mu2e experiment from target to detector is shown in Fig. 2. Particles produced in the target are directed along the production solenoid and into the bent solenoid transport solenoid, which selects beam that continues into the Al stopping target within the detector solenoid. The transport is designed to accept low energy μ, mostly produced from π decay within the solenoids. The bent solenoid and associated collimators are tuned to accept ~0—100 MeV/c μ; stopped μ's are obtained from ~0—50 MeV/c μ. Figure 3 shows the momentum distribution of muons reaching the detector solenoid, including the population of those that are stopped within the target. Peak stopped μ's are actually at 35-40 MeV/c (6—7 MeV kinetic energy). If we could put more of the beam from the ~50—100 MeV/c into the < ~ 40 MeV/c region, the useable mu2e beam would increase.

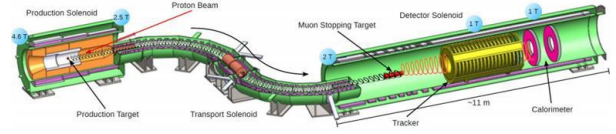

Figure 2: Overview of the mu2e experiment from production solenoid to transport solenoid to Detector Solenoid.

The transport solenoid (TS) consist of two bent solenoids with a short transition solenoid connecting them; collimators are in the transition solenoid. The equations of motion in a bent solenoid are:

$x'' = h + by' \qquad y'' = h - bx'$

where $b = B_o/B\rho$ and $h = 1/\rho_0$, where $\rho_0$ is the bending radius and $B_0$ is the solenoidal field. In mu2e the bending radius is ~ 3 m. The magnetic field is $B_0 = 2.4$T and $B\rho$ (T-m) = p (GeV/c)/0.3. The total bend of each bent solenoid is 90°, but with opposite signs. The solutions of these equations are:

$$y(s) = C_1 - \frac{h}{b}s + \frac{h}{b^2}\sin(bs) + C_2 \frac{1}{b}\sin(bs) + C_4 \frac{1}{b}(\cos(bs)-1)$$

$$x(s) = C_3 + \frac{h}{b^2}(1-\cos(bs)) + C_2 \frac{1}{b}(1-\cos(bs)) + C_4 \frac{1}{b}\sin(bs)$$

$C_1 = y(0), C_2 = y'(0), C_3 = x(0), C_4 = x'(0)$. Starting at 0 in all coordinates, we obtain the solution:

$$y(s) = -\frac{h}{b}s + \frac{h}{b^2}\sin(bs); \quad x(s) = \frac{h}{b^2}(1-\cos(bs))$$

At the end of the first bent solenoid, entering the transition, s=4.6m. At that point the vertical position is linear with momentum p: y(p) = ~0.213 (p (MeV/c)/100) m for negative muons. (opposite sign for positive muons). The oscillation amplitude $h/b^2$ is a measure of beam oscillation within the focusing solenoid fields and is relatively small (~0.16cm for p=50MeV/c).

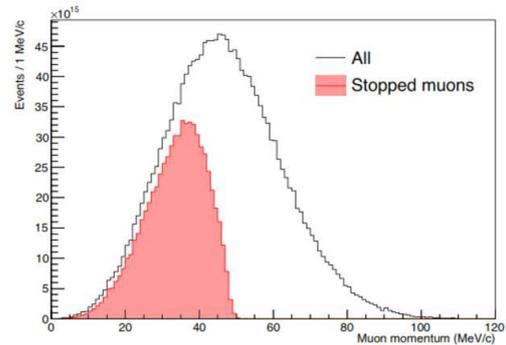

Figure 3: Simulated muons that reach the detector solenoid. Stopped muons that can provide mu2e candidate events are displayed in red. From ref. 1.

Muons in the ~40—100 MeV/c momentum range can have their momenta moved toward the ~40 MeV/c desired momentum by passing through an absorber with a thickness corresponding to the desired momentum loss. This can be done by using a wedge with a thickness that depends on vertical position. The requirement would be that the thickness as a function of y be such that $P_\mu(y)$ is reduced to ~40 MeV/c. This implies zero thickness for y < 8.5cm. Momentum loss in material is strongly momentum-related. From the Bethe-Bloch formula:

$$p(s) \cong p_1 - \int_0^s 0.3071 \frac{Z}{A} \rho \left( \frac{(p^2 + m_\mu^2 c^2)}{p^2} \log\left( \frac{2p^2 c^2 m_e c^2}{I_e (m_\mu c^2)^2} \right) - 1 \right) \frac{\sqrt{p^2 + m_\mu^2 c^2}}{p} ds$$

We can choose the thickness as a function of position y by requiring that the integrated energy loss be equal to the amount necessary to reduce the momentum to ~ 40 MeV/c.

$$\text{range}(40, p_1) = \int_{40}^{p_1} \frac{1}{\frac{dp}{ds}(p)} dp$$

Combining this with the relationship $y(p_1) = \sim 0.213 \, (p_1 \text{(MeV/c)}/100)$ m, obtains the desired thickness as a function of y.

The choice of a wedge material is dependent on practical considerations that are not fully explored within this note. The material should be a relatively low-Z material to minimize multiple scattering and must be mechanically compatible with the transport solenoid vacuum pipe. For an initial estimate we consider polyethylene (~$CH_2$), with properties of Z/A = 0.57, ρ=0.94, $I_e$= 57.4 eV, $X_0$ = 47.45 cm. This is a low cost material that is easily machined, and has actually been used to produce a wedge for the MICE experiment, (see Fig. 4.).

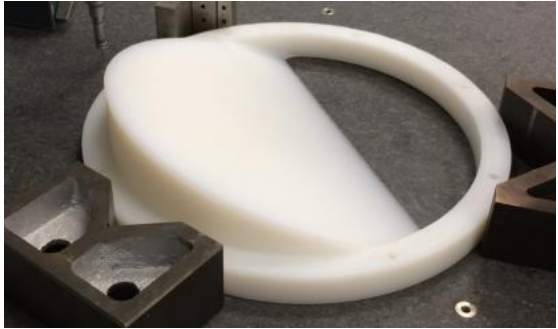

Figure 4: Polyethylene wedge piece machined for use in the MICE experiment. Wedge for mu2e would have similar dimensions. (Inner radius is ~20cm.)

Evaluation of the range equation obtains an absorber with zero thickness for y < 8.5 cm. For y =0.213m ($P_\mu$=100 MeV/c), a thickness of 7.22cm is required, for y=24cm, 9.15 cm and for y=15 cm, 1.61cm. A numerical solution and fitting within Mathematica obtains a thickness of a polyethelene absorber given by:

$w = 0$ for $y < 8.5 cm$

$w \cong 0.133(y - 8.5) + 0.0296(y - 8.5)^2$ cm for $y > 8.5 cm$

This nonlinear profile is roughly parabolic. [Fig. 5] The TS has an inner radius of ~25cm, so the maximum thickness would be ~10cm. For denser materials (Be or $B_4C$) the thickness would be ~5—7 cm. This maximum thickness could be reduced by reducing the maximum matched $P_\mu$ to be reduced. (3—5 cm for 100 MeV/c).

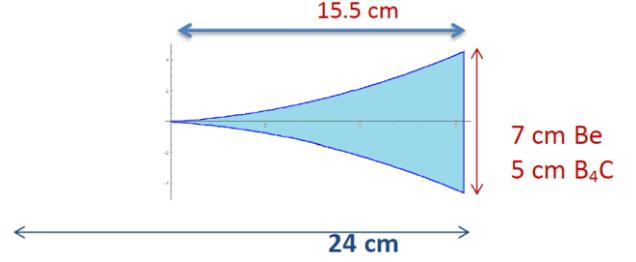

Figure 5: Cross-section of a wedge to be inserted at the high dispersion point of the mu2e transport solenoid. The wedge is offset from the center line by ~8.5cm so that low-energy μ's pass without touching the wedge; higher-momentum μ's lose energy toward $P_\mu$ ~40 MeV/c.

## APPLICATION TO G-2

The g-2 ring has a very small momentum acceptance for 3.1 GeV/c μ's (δP/P is ~0.1%). The beam transport into the g-2 ring (which includes the Debuncher ring has a much larger acceptance of ~1%. Reduction of that momentum spread before injection into the ring could increase the number of accepted μ's. This would require a wedge absorber at a point of the transport with non-zero dispersion. While larger dispersion is desirable, the transport as presently designed has dispersion η < ~1m. At that dispersion an offset of 0.001 δP/P is 1mm and to change momenta to the baseline requires a momentum loss of 3.1MeV/c.

We would like the absorber thickness also to be less than a few mm, which means using a dense material (large dE/dx). We note that dE/ds at 3.1 GeV/c is ~20% larger than minimizing ionizing. We also want minimal multiple scattering, which would mean low-Z materials. High-density low-Z materials could include: Be (dE/ds=3.8 MeV/cm), Boron Carbide $B_4C$ (dE/ds = 5.5 MeV/cm) or diamond (dE/ds = ~8.4 MeV/cm). Polyethelene would have dE/ds = 2.56 MeV/cm. Since the absorbers are relatively thin, higher Z materials can be considered, such as Nickel (dE/ds =18.2 MeV/cm) or Iridium (dE/ds =36 MeV/cm). For reduction of energy offsets to zero (with small transverse emittance beam size) we require: $2 \frac{dE}{ds} \eta \tan\left(\frac{\theta}{2}\right) = E_\mu$ .

For η=1m, θ = 161º, 152.5º, 141º, 123 º, 81º, 47º for polyethelene, Be, $B_4C$, diamond, Ni, and Ir, respectively. For η=0.65m, θ = 168º, 162 º, 154º, 141º, 105º, 67º for polyethelene, Be, $B_4C$, diamond, Ni, and Ir, respectively.

The transverse and longitudinal acceptances of the ring are limited by the ring aperture, which has an r = 4.5 cm radius and the betatron functions ($\beta_x$ = 8m, $\beta_y$ = 18.4m, η=8.2m). These are given by:

$$A_x = \frac{r^2}{\beta_x}, A_y = \frac{r^2}{\beta_y}$$, which is 0.00025m for $A_x$ and

0.00011 m for $A_y$. This could be rewritten as an rms normalized emittance acceptance by multiplying by βγ = 30 for $P_\mu$ = 3.1 GeV μ's and dividing by 6 ($\varepsilon_{rms}$ =A/6), obtaining $\varepsilon_x$ = 0.00125m, $\varepsilon_y$ = 0.0005m. The transverse emittance increase induced by the absorber must be significantly less than these aperture cuts. (Stricter aperture cuts are imposed by the injection optics.)

The increase in rms normalized emittance caused by multiple scattering in the absorber can be estimated by

$$\delta\tilde{\varepsilon}_N = \beta_t \frac{E_s^2}{2\beta^3\gamma(mc^2)^2} \frac{\Delta w}{L_R},$$

where $E_s$ = 13.6 MeV, $\beta_t$ is the transverse beta function at the wedge, $L_R$ is the material radiation length, $\Delta w$ is the length of the absorber at central energy loss, $mc^2$ = 105.66 MeV, and $\beta$, $\gamma$ are the relativistic kinematic factors ($\beta$ ~1, $\gamma$ ~30). If we require the energy loss at central momentum to be 2×3.1MeV, set $\beta_t$=10m (typical for a transport line), then $\delta\varepsilon_N$ = {0.00013, 0.00015, 0.00017, 0.0008, 0.0016} m for Be, B$_4$C, C, Ni, Ir, respectively. The geometric emittance change is $\delta\varepsilon_N$ /29.3 or {4.4, 5, 6, 27, 53} mm-mrad (geometric). While the low-Z materials may have tolerably small emittance increases, the higher-Z materials lead to emittance increases that are large when compared to the acceptances. They would not be recommended, unless $\beta_t$ is significantly reduced. The application to g-2 appears to require low-Z materials, with large angle absorbers. Figure 6 shows a schematic view of the muon beam passing through a large angle absorber.

Figure 7 displays betatron functions for the transport into the g-2 ring. The transport to the ring has a horizontal dispersion of ~0.65m where $\beta_x$ = ~2m and $\beta_y$ = ~7m, and this might an appropriate location for the wedge.

However the transverse beam size due to emittance is not small, even with $\beta_x$ = ~2m. With an rms emittance of 12 mm-mrad, the rms beam size $\sigma_x$ is 5mm. A large transfer in emittance can occur if $\eta\delta_0 \gg \sigma_x$, where $\delta_0$ is the initial beam width. At $\delta_0$ =0.01 and $\eta$=0.65m, $\eta\delta_0$=6.5 mm which is similar to the emittance beam size. Within the linear wedge model, the $\delta'$ to obtain minimal $\delta p_{after}$ is reduced from $1/\eta$ to $1/\eta \times ((\eta\delta_0)^2/((\eta\delta_0)^2 + \sigma_x^2))$, and $\delta p_{after}$ is given by

$$\delta p_{after} = \delta_o \left(1 - \frac{(\eta\delta_0)^2}{(\eta\delta_0)^2 + \sigma_x^2}\right)^{1/2},$$

which is ~$0.61\delta_0$ at the above parameters. Optimum $\delta'$ is reduced by a factor of 0.63 from the zero beam size limit, which reduces the desired wedge angles substantially. For this optimum $\delta'$, the dispersion is matched to zero exiting the wedge; a lattice exploiting this feature could have significant advantages.

A larger dispersion function at the wedge would be more desirable, of course.

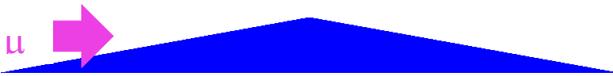

Figure 6: Schematic view of a muon beam incident on a large angle (160º) absorber.

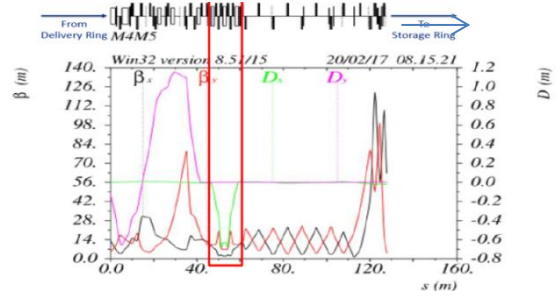

Figure 7: Betatron functions for the transport leading into the g-2 ring. Test wedges are inserted at high dispersion within the highlighted region in simulation.

## SIMULATION RESULTS

We have initiated simulations of the g-2 case. In these simulations, Beam was tracked using G4Beamline through the transport from the delivery ring into the g-2 ring. Wedges of various materials and dimensions were inserted in the high dispersion region and optimized for providing the most beam within a ±0.1% $\delta p/p$ acceptance window through the transport.

As may be expected from the above discussion, the optimum wedge is a low-Z material (poly (~$C_2H_4$) or LiH) with a shallow angle (~150-160º), and increases the beam within the acceptance by ~30%. (see fig. 8) The wedge increases transverse emittance, and with the mismatched optics the larger amplitude particles were lost in the transport; the 30% net improvement included the losses.

The result is considered to be enough of an improvement to encourage further development, including further simulation and design and construction of moveable physical wedge inserts in the g-2 transport line at the high dispersion point.

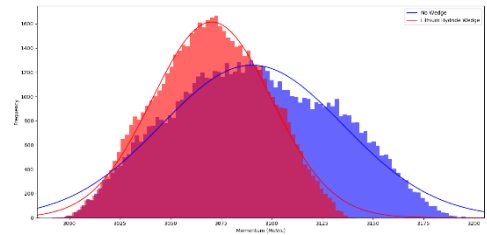

Figure 8: Momentum distribution of beam reaching the g-2 ring without (blue) and with (orange) a wedge. While total number of muons is reduced, the momentum width is reduced and beam within ±0.1% is increased by ~30%.

In the initial evaluations the beam optics was not rematched following the wedge, and a full simulation with matching into the g-2 ring and storage has not yet been completed. This must be done in the near future. Proper matching could increase acceptance significantly. A modified optics with larger dispersion at the wedge could also greatly improve acceptance.

## ACKNOWLEDGEMENT

We thank M. Syphers for helpful and encouraging discussions.